\documentclass[sigconf]{acmart}
\copyrightyear{2025}
\acmYear{2025}
\setcopyright{rightsretained}
\acmConference[SIGMOD-Companion '25]{Companion of the 2025 International
Conference on Management of Data}{June 22--27, 2025}{Berlin, Germany}
\acmBooktitle{Companion of the 2025 International Conference on Management
of Data (SIGMOD-Companion '25), June 22--27, 2025, Berlin,
Germany}
\acmISBN{979-8-4007-1564-8/2025/06}
\acmDOI{10.1145/3722212.3724480}

\usepackage{balance}
\usepackage{graphicx}
\usepackage{color}
\usepackage{listings}
\usepackage{subcaption}
\usepackage{enumitem}
\usepackage{xspace}
\usepackage{xcolor}
\usepackage{colortbl}
\usepackage{url}
\usepackage{hyperref}
\usepackage{cleveref}

\newcommand{\stitle}[1]{\smallskip\noindent{\textbf{#1}}\xspace}

\begin{document}
\setlist{leftmargin=*,itemsep=0pt}

\newcommand\vldbdoi{XX.XX/XXX.XX}
\newcommand\vldbpages{XXX-XXX}
\newcommand\vldbvolume{18}
\newcommand\vldbissue{14}
\newcommand\vldbyear{2024}
\newcommand\vldbauthors{\authors}
\newcommand\vldbtitle{\shorttitle} 
\newcommand\vldbavailabilityurl{}
\newcommand\vldbpagestyle{empty} 

\newenvironment{myenumerate}{%
\begin{enumerate}[leftmargin=1em, itemsep=.1em, parsep=.1em, topsep=.1em,
    partopsep=.1em]}
{\end{enumerate}}


%
%



\setcounter{figure}{0}
\setcounter{section}{0}

\title{Where Does Academic Database Research Go From Here?}

\author{Eugene Wu}
\email{ewu@cs.columbia.edu}
\affiliation{%
  \institution{Columbia University}
  \city{New York}
  \state{NY}
  \country{USA}
}

\author{Raul Castro Fernandez}
\email{raulcf@uchicago.edu}
\affiliation{%
  \institution{The University of Chicago}
  \city{Chicago}
  \state{IL}
  \country{USA}
}

\date{}
\renewcommand{\shortauthors}{Eugene Wu \& Raul Castro Fernandez}


\begin{abstract}
    An open forum to discuss and debate the future of database research in the context of industry, other research communities, and AI.
\end{abstract}



\maketitle

\section{Introduction}
\label{sec:intro}

We can informally define an academic community as a group that jointly contributes to advancing knowledge within a specific discipline, with shared research questions and methodologies.  Communities are brought together due to the importance and impact – broadly defined – of the questions that the communities pursue.  

The world, of course, is not sentimental.  The fortunes of a community ebb and wane based on its competitive advantage as compared to the rest of the world.   Historically, alchemy faded in prominence and was replaced by more effective methods, namely chemistry and the scientific method.  Similarly, humoralism was replaced by evidence-based medicine, and behaviorism was replaced by cognitive psychology and neuroscience. 
In engineering, cybernetics grew out of war efforts and appeared poised to take over the world, but ultimately evolved into the pragmatic discipline of control theory.  In computer science, ``machine learning'' research as repeatedly evolved from the 50s (logic-based), the 70s (expert systems), to today's deep neural networks and LLMs that have taken center stage due to the massive potential that automated intelligence can bring.    

Closer to our hearts, database systems have grown in prominence over the past 50 years due to the demands of accounting, logistics, the Internet, and digitization. Yet, the data management success does not guarantee future success; that is up to us---the community---to chart our place in the AI era.


A recent SIGMOD blog post~\cite{wublog} reflected on the state of academic research from this perspective of competitive advantage as compared to startups, industry, and other research communities.   It argued that much of the success of the data management community is tied to the commercial success of relational database management systems (RDBMS). Yet, this success has transformed RDBMSes into a commodity that sits on the plateau of the technology S-curve, and continued investment provides marginal gains that cannot compete with industry.    At the same time, the article asked what advantage academic database research has in AI.

If this argument is true, it shakes the foundation of our academic community.  We must answer what our role is and how the problems we solve are relevant not only to the researchers associated with our community but also to the outside world.   On the other hand, many members of the community have objected to this characterization, and argue that our community is healthy, relevant, and simply poorly marketed to the outside world.    

This naturally sets the stage for a debate, and it is in this context that we have organized a panel of six members of the database community to debate the direction of the academic database community and where we should be going to maintain a competitive edge.    Our goal is to identify directions and tangible action items to help ensure our community continues to be relevant in the age of AI and beyond.  The panel will be structured around the following themes:
What is the competitive advantage of the academic database community over the next decades? 
What is needed to keep it relevant? 
What specific interventions can we implement? 
What is our place in the current AI revolution?

\smallskip\noindent\textbf{This Document:}  The published ACM SIGMOD and VLDB panel proposal documents should be considered as snapshots several months before the conferences.   An up-to-date version of this document with comments from across the community can be found on ArXiv under the same title.

\subsection{Self-Reflection Through the Ages}
The data management community has a history of self-reflection.  Roughly every 5 years, a select group of senior academics convenes a closed-door meeting to discuss trends, opportunities, and challenges for the database community and issues a report~\cite{bernstein1998asilomar,abadi2016beckman,silberschatz1996strategic,abadi2020seattle}.   In contrast, this panel is an open forum for the SIGMOD community to participate in the discussion. Members of the community have spoken out on threats to the community, including Dewitt’s 1995 keynote~\cite{dewitt} and Stonebraker’s 2018 Ten Fears talk~\cite{stonebraker}.

\section{VLDB 2025 Panel Composition and Format}

\subsection{Composition and Format}
For VLDB in August 2025, we again solicited additional positions from 5 members of the community (in addition to those we previously solicited and talked to at SIGMOD.   In contrast to the SIGMOD panel, which was composed of more senior researchers across data management topics, geographies, and roles, the VLDB panel varied in seniority, with panelists that range from PhDs to senior members.   We believe that this diverse mix of perspectives will unlock new insights and answers to the questions above.

Each panelist will present a 1-minute thesis, followed by a 1.5-hour moderated panel discussion with interactive inputs from the audience.   The panelists are, in no particular order:
\begin{itemize}
    \item \textbf{Shreya Shankar}, 5th-year PhD student in EECS at UC Berkeley.
    \item \textbf{Natacha Crooks}, Assistant Professor of EECS at UC Berkeley.
    \item \textbf{Jiannan Wang}, Associate Professor at Simon Fraser University
    \item \textbf{Gustavo Alonso}, Professor of Computer Science at ETH Zurich.
    \item \textbf{Divesh Srivastava}, head of Database Research at AT\&T, President of the VLDB endowment, Co-chair of the ACM Publications Board, ACM Fellow.
 \end{itemize}

\subsection{Initial Panelist Takes}

\subsubsection*{Shreya Shankar}
One thing very special about our community is how we span the full stack---from building intuitive interfaces to tuning hardware---all while staying grounded in a common goal: helping people unlock value from their data. This versatility is our strength. As long as we remain intellectually diverse and receptive to new ideas, I think our field will continue to flourish. We don't need to measure ourselves against other communities (e.g., AI); the steady growth in students pursuing data management PhDs and applying for data management faculty positions speaks for itself.

Yes, we experience waves of hype, and it may be challenging to do novel research in areas after they've achieved commercial success. But history consistently shows that new tech frequently emerges, presenting fresh opportunities for us to design efficient, usable systems. Also, even after the hype fades in an area, some problems remain unsolved. So, there will always be space for us to apply our data management expertise, perhaps reimagining aspects of our approach along the way---and that is precisely what keeps our field dynamic and relevant.

\subsubsection*{Gustavo Alonso.} The notion that research in data management and database techniques is obsolete and has been superseded by something else is a recurring theme in the community. We have gone through several crises that all proved to be unfounded (the web, search engines, the cloud, eventual consistency, key value stores, AI/ML, LLMs, etc.). This is highly surprising and somewhat amusing once one notices the pattern. New generations of researchers bring fresh ideas and are closer to recent developments but have a tendency to overestimate the novelty and value of new trends. Senior researchers often claim everything was done years ago and there is little new and original work being pursued, which leads them to underestimate the potential of those new trends. When one half of the community gets easily excited by outside topics and the other half sees little novelty in the work being done in the area, we are bound to be constantly questioning our purpose. Wild marketing, hype, and outrageously misleading claims from industry do the rest.

The cloud is a great example of this process. The emergence of the cloud led to many outlandish notions: transactional consistency was irrelevant, SQL had no purpose, schemas and data formats only got on the way, etc. Twenty plus years later, strong consistency is still a must and being enforced even at very large scales, SQL is a pervasive as ever, and there is a huge amount of work being done in all sorts of data formats and schema representations. Why this happened is easy to explain in hindsight, but it is a pattern that repeats itself with new technologies. The cloud represented a major shift in architecture, embracing the scale out systems used in search engines and web servers that worked very well for embarrassingly parallel workloads. At the beginning, it was not obvious how to transfer the mechanisms available in conventional relational engines to the cloud or even that such engines were still relevant. This led to the claims that database research was dead and to the usual conference panels complaining about the attention having moved somewhere else. It also triggered a defensive reaction by parts of the database community with claims that innovation taking place in the cloud was clearly inferior to traditional engines. As usual, the two sides were both right and wrong. It took time to adapt and evolve data management techniques to the cloud. In fact, about two decades. But we are there, with data management concepts being as relevant as ever and with, e.g., data analytics still making up a big part of the workloads and revenue of cloud providers. At the same time, the cloud required to rethink many established assumptions and designs (in particular, to break away from the notion that data processing happened on a database engine), which led to a great deal of innovation, as well as new ideas and products. Far from killing database research, the cloud introduced new dynamics and challenges that we are still trying to address with the practical relevance of data management having expanded rather been diminished.

Now we are confronted with AI, ML, and LLMs and the world is seemingly caving in on us once more. Wild claims are being made about AI: everything -even people- is going to be replaced by LLMs, nobody will use SQL because queries will be formulated in natural languages, performance tuning will be automatically taken care of by some AI agent, perfect engines will be miraculously developed  by ML, query optimization will be a thing of the past, relational engines will not be needed since an LLM will answer all possible queries instead, etc. We are back to a world where data management seems to be less relevant and the focus of attention has shifted somewhere else. Yet, all these systems are based on data, on huge amounts of data that will need to be managed much better than it is today. It is a safe bet that, over time, the techniques and research being done in the database community will be a crucial component to the AI revolution. We are not there yet and will take us many years to figure it out but we have a lot to contribute. AI, ML, and LLMs are a great opportunity for the database research community if we focus on our strengths and can avoid the hubris around the AI hype. These are data systems and, like many such systems in the past, starting with the relational model, they need to be made more efficient, their answers more precise, data needs to be kept consistent, interfaces should be more declarative to allow for better optimizations, better data formats and representations need to be found, processing needs to be adapted to the new hardware platforms,  etc. It is a challenge but it will keep us active and relevant for many years to come.

\section{SIGMOD 2025 Panel}

\subsection{Composition and Format}
For SIGMOD in June 2025, we solicited positions from 15-20 members of the community across different geographies, positions, seniority, and research topics.  We then chose six panelists that provided balance in terms of academia and industry, applied and theoretical research, and that represented positions that vary from ``stay the course'' to ``AGI will make everything irrelevant.''
Each panelist will present a 1-minute thesis, followed by a 1.5-hour moderated panel discussion with interactive inputs from the audience.   The panelists are, in no particular order:
\begin{itemize}
    \item \textbf{Dan Suciu}, Microsoft Endowed Professor in Computer Science and Engineering at the University of Washington.
    \item \textbf{Sihem Amer-Yahia}, Research Director, CNRS, LIG; Vice President of VLDB Endowment.
    \item \textbf{Yannis Ioannidis}, Professor of Informatics \& Telecom at the University of Athens, President of the ACM.
    \item \textbf{Ippokratis Pandis}, VP/Distinguished Engineer at AWS.
    \item \textbf{Jens Dittrich}, Professor of Computer Science at Saarland University; 3x CIDR gong show winner.
 \end{itemize}

\subsection{Panel Overview}

As organizers, we aimed to foster a bottom-up discussion on the evolving role and future direction of academic database research. Motivated by seismic technological and budgetary shifts, particularly the rise of AI, we explored the comparative advantage of the academic database community. This advantage is critical because it dictates our ability to recruit top students, ensure their success after graduation, and maintain our relevance in a rapidly changing technology and economic landscape. We specifically emphasized that "we" refers to academic database research, contextualizing the topic around recent shifts in government funding (\$4B NSF expected budget for 2026) compared to the rapid growth of private investment in AI (e.g., >\$100B VC funding in 2024, >\$50B in Q1 2025), as funding directly influences the talent landscape.

The feedback for the panel and its structure was as varied as the recommendations themselves. Many approached us afterward to share their excitement in the panel’s discourse and for surfacing their worries and challenges. Others felt the existence of such a panel was yet another in a "sorry tradition of needless negativity"\footnote{\url{https://jhellerstein.github.io/blog/sigmod-optimism/}} for a vibrant and thriving research community.   With a community as big as ours, we are grateful to receive these diverse opinions.

Matthew Butrovich and Lampros Flokas kindly took notes.  The quoted text below is pulled from these notes, so we do not attribute them to specific individuals.

\subsection{Panel Summary}

The discussion revolved around two primary topics: the competitive advantage of the academic database research community and actionable strategies for its future.

\subsubsection{Our Competitive Advantage}
The first topic focused on the academic database community’s competitive advantage as compared to other disciplines, industry, and startups.

\stitle{Data Principles and AI:} Our community's core database principles, such as "independence between physical and logical" and "declarativeness”, and “automatic scalability hold lasting value. The CTO of Alibaba Cloud demonstrated this by applying query optimization principles to rebuild their pipeline for training QWEN 3, significantly reducing costs. We believe "all AI problems boil to data," and our deep systems knowledge offers unique insights into challenges like data provenance, security, and novel data abstractions for AI workloads. A panelist noted the "bug" of companies needing hundreds of engineers for each of their separate "AI stack" and "data stack" teams, highlighting a significant opportunity for unified solutions rooted in database thinking.

While some external communities might prioritize immediate problem-solving and "accuracy" over these foundational principles, WE must effectively communicate their long-term benefits. "The way we interact with other communities is problematic" and "we don't communicate what we have in the right away."

\stitle{Navigating the Modern Data Landscape Beyond the Traditional DBMS:} “DBMSes are fast enough” for most applications, and our community's expertise is vital for new data problems beyond the classic RDBMS. For instance, many performance problems are due to the ORM and never arise at the DBMS, while societal-level issues like fake news with an ongoing wave of AI-generated content (“AI slop”) is an information integrity problem. A panelist pointed out that a MacBook can comfortably run TPC-H scale factor 1000: “small data” is enough for most applications and challenges lie in programmability, interoperability, and usability. Conversely, an audience member argued that improving the performance of RDBMSes can also be valuable as throughput increases may unlock new types of applications.

In the AI world, "solutions are crappy when you" combine diverse workloads like vectors, keywords, and relational queries in commercial systems. The platform for building data management systems is "radically different" at "cloud scale," involving "smart NICs, disaggregated memory that comes and goes, [and] GPU spending." These new hardware environments present unique opportunities for our deep systems knowledge.

\stitle{Academic Research Identity:} Balancing Curiosity, Impact, and Resources: Academia holds a unique position to pursue "curiosity driven research," which contrasts with industry's often shorter-term, product-focused cycles. This allows for "academic play" that breaks free from the need to make numbers go up. Panelists observed that "industry thinks 3/4 quarters ahead" while "Academia thinks 5 years ahead." Despite industry's "muscle" (e.g., 600 people on projects compared to "2 students for 2 years" in academia), academia retains the freedom to explore complex, foundational questions without immediate commercial constraints. This implies that academia should not directly compete with industry; as Surajit mentioned in the Tuesday panel, researchers should look for high upside when choosing a problem.

\subsubsection{What the Academic Community Should Do?}

The second half of the panel shifted towards actionable takeaways.

\stitle{Bring the Data (Researcher) To the Problem:} “The way we interact with other communities is problematic” and we expect other communities to come to us. Instead, 1/3 of research sessions in other conferences have ‘data’ in their titles and “these communities are not going to come to SIGMOD”. A key recommendation was to proactively engage with other academic disciplines. The panel stressed the need to "go out" to these communities, and that "inter-disciplinary research post tenure" could be "tremendously fulfilling". One suggestion was to create "Data + X" workshops in the X communities that are run in a "true discussion mode not as a paper presentation".

\stitle{Win the Heart of Data Stakeholders:} Despite the lasting value our core database principles, such as "independence between physical and logical" and "declarativeness," hold, we should not waste our efforts convincing other CS disciplines that these matter. We should direct our efforts toward winning the hearts of data stakeholders. How do we do that? By showing how good we are at treating everything as data and building the data backbone of their needs.  

\stitle{Improve Dissemination and Communication:} There was a strong call to make academic research more accessible and understandable, as "nobody reads our papers, but they want to talk to someone who reads our papers". It was suggested that papers are "extremely hard to read," and the community "let[s] others pull the knowledge". Recommendations included writing "better tutorials" and producing "YouTube videos", and reducing SIGMOD papers to "8 pages" accompanied by "a clean and fun youtube video".

\stitle{Foster Industry-Academia Collaboration:} Encouraging students to spend time in industry was highlighted as a way to expose academic research to real-world challenges and bring practical problems back. It was noted that "OpenAI doesn't come here, but they bought a database" company, Rockset. Their “head of engineering is from Almaden. Send students into industry for a bit after teaching them systems.”

\stitle{Rethink Conferences and Publication Norms:} The panel discussed the need to innovate scholarly communication and conference formats. SIGMOD is a leader in pushing scholarly communication forward (e.g., we introduced reproducibility efforts), and there are ongoing discussions at the SIGMOD and ACM levels about reorganizing conference formats to better support "conferring" over listening to "canned presentations".   Continuing the idea of academic play, a panelist remarked that our reviewing process may over-index on work that must be faster than some baseline, and inadvertently prune the early seeds of ideas that may not lead to impressive results now but open up impactful research directions in the long run.

\stitle{Strategically Focus on Foundational Principles in New Contexts:} Panelists argued that we need not "become experts in LLMs" or other domains. Instead, our strength lies in forming "multi-disciplinary teams" where each contributes their core expertise. Specifically, for AI, "LLMs are a race to 0. It's a tool." Instead, "the application and the data is interesting and where the money is". The community's advantage lies in knowing "complexities of modern technology like GPUs" and continually learning.

\subsubsection{Closing Comments}

The panelists were asked to imagine they were “president of database research” and could unilaterally force one change.

\begin{itemize}
    \item “Go out and create workshops that are ‘data + X’ or ‘data 4 X’, have them run in a true discussion mode not as a paper presentation with two questions. Interactive panels and a few keynotes by experts. Run interdisciplinary workshops and everyone is forced to go to at least one of them”
    \item “Force everybody to send students to industry for a bit and bring problems back. security, provenance, data integration. Force a change of mind that academia is the holy land and let students be more open to industry.”
    \item “Impose that SIGMOD papers be reduced to 8 pages. accompanied by a clean and fun youtube video.”
    \item  “Help people build real applications and make it open source.”
\end{itemize}

\subsection{Reflections From the Organizers}

Overall, we came away from the panel excited that our community has so much impact, and possesses the data-centric tools that the world needs!    However, our future impact hinges on proactively engaging with a rapidly evolving landscape of applications, evangelizing our expertise to other communities, and strategically collaborating across domains.   All great motivations for academic research!

The positive and insightful discussion ultimately led to recommendations that exhibit inherent trade-offs, rather than a single, clear set of recommendations:

\begin{itemize}
    \item \stitle{Is Database Research Art?}  We should leverage academia’s unique capacity to pursue curiosity-driven research (academic play), but a real and present budget crisis exists. How do we balance these two imperatives? How can academic research secure sustainable funding for foundational, long-term problems that industry cannot or will not pursue, without becoming irrelevant to the capitalist marketplace that attracts top talent and resources?   It reminds us of artists that strive to pursue purity in expression but face economic realities.
    
    \item \stitle{Impactful Research vs. Market Dominance:} We should do impactful research that is neither too short-term nor too long-term, yet offers high upside. However, the relational database market completely dwarfs any alternative data applications and domains. How do we balance the dominance of existing database technologies with the need to explore new paradigms that may not have immediate commercial application but might offer long-term, high-upside impact?
    
    \item \stitle{Go Big and Stay Real:} We should talk to other domains and experts to be “trans-disciplinary.” But how do we initiate such projects effectively? What specific models for "multi-disciplinary teams" can integrate deep database knowledge into diverse fields (e.g., AI, quantum computing, social sciences) to solve "data problems" without sacrificing the rigor and identity of database research?
    
    \item \stitle{Engineering vs Science:} We encourage students to engage with industry to gain real-world experience, but we also aim to conduct long-term, ambitious scientific research. How do we ensure that industry exposure enriches, rather than detracts from, fundamental scientific inquiry? On the other hand, data management is an applied field, so does it matter?
\end{itemize}

\noindent In either case, we do feel that problems should be paired with solutions. We plan to re-structure the VLDB 2025 panel in late August to delve deeper into concrete trade-offs in the above recommendations.

\subsection{Initial Panelist Positions}

The following were the panelist positions solicited before the event.

\subsubsection*{Dan Suciu}
Our community has a lot to offer to computer science in general.  But
we are not doing a great job at advertising our techniques beyond the
success of relational databases. (Perhaps we can debate what we should
do about this.)  Yet, the people who need us, they know where to find
us: the top database systems PhDs are hired with insane compensation
packages, researchers from other areas increasingly borrow techniques
developed by the DB community, and (to my pleasant surprise) even
database theory has seen increased interest recently.

\subsubsection*{Sihem Amer-Yahia}
My position is to reach out to colleagues in other disciplines who find themselves solving data-intensive problems the best they can, and help them better understand the mapping between their needs and data and the multiplicity of approaches out there. Why us? because we care about data models and because our ability to separate the what from the how and the why will help them better apprehend their questions. In other words, we are the best positioned to structure the data science that powers multi-, inter-, cross-, and trans-disciplinary research (Figure below). 
{\it Can we rethink database design to enable these different ways of conducting science?  }

\noindent \textbf{Multi-disciplinary research} involves several academic disciplines working in parallel on a common goal, yet following their individual disciplinary precepts and approaches. As an example, large-scale hypothesis testing requires to understand statistics and algorithms separately. The key point is that scientists exchange knowledge, but do not aim to cross subject boundaries to create new, integrated knowledge and theory. 

    \includegraphics[width=0.8\columnwidth]{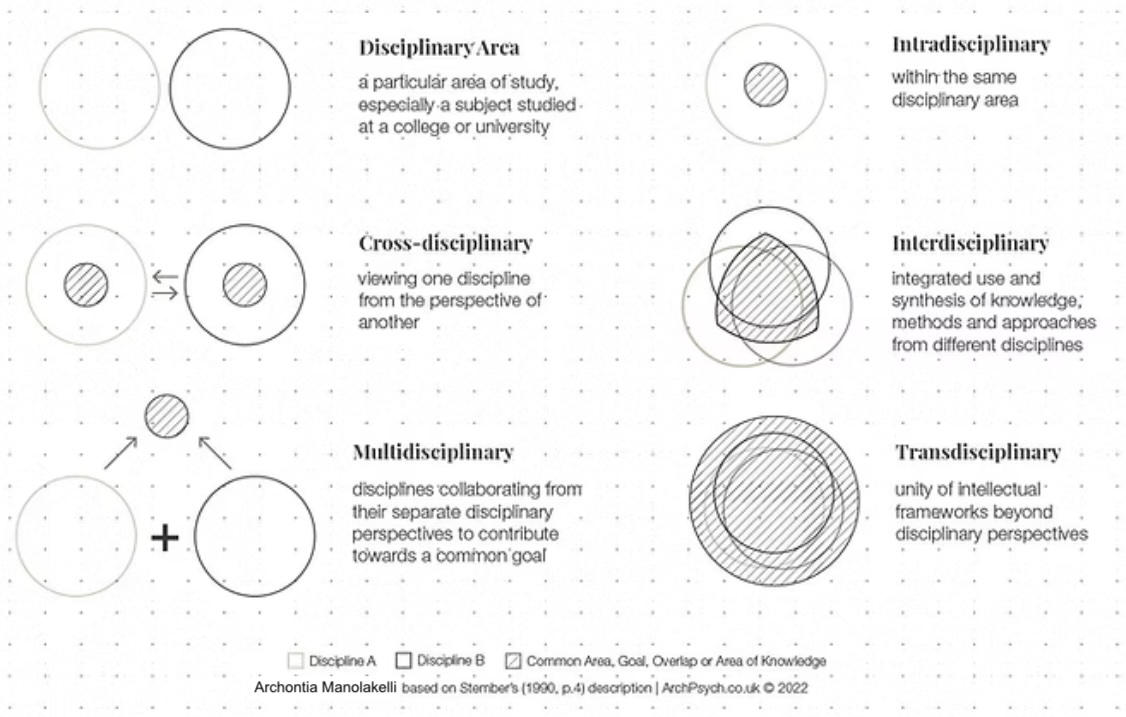}

\noindent In \textbf{inter-disciplinary research}, boundaries between disciplines are crossed. Social Computing, for instance, is a new body of knowledge that resulted from integrating data management, web science, machine learning, and social science theories. The key point is that scientists pool their approaches and modify them so that they are better suited to the questions at hand. 

\noindent \textbf{Cross-disciplinary research} allows researchers to combine different perspectives and methods borrowed from other disciplines, potentially leading to more innovative and holistic solutions to a shared problem. For example, viewing database query processing from the perspective of information retrieval led us to top-k query processing. The key point is that scientists learn to view one discipline from the perspective of another. 

\noindent \textbf{Trans-disciplinary research} transcends academic and work realms and involves collaboration between researchers, practitioners, policy makers and communities, with the aim of addressing real-world problems and creating practical solutions. A trans-disciplinary activity is usually values-centered and firmly situated in the present, its participants varied and diverse. Take as an example studying algorithmic fairness that brings together DB, ML and Law because of everyone’s interest in the same cause. The key point is that the bodies of knowledge from each of the participatory scientists are blended to form new methods and practices. 

\subsubsection*{Yannis Ioannidis}
[no comments submitted]

\subsubsection*{Ippokratis Pandis}
No position!


\subsubsection*{Jens Dittrich}

 This panel might be a very transient thing, once AI takes over the entire world and does research itself (which is already happening to some degree), then what is the point? (Am I kidding? No.)
 Meanwhile:

\begin{myenumerate}

    \item I have some doubts whether the AI/ML folks need us in the long run. Sure, declarativity and scalability. Yes, we could help. But the solutions they need are relatively simple, not really researchy, feels more like using/teaching best practices from DB. (also see 4.) But maybe I am wrong.
    
    \item As researchers we live in this research bubble where we can spend the entire day thinking about our database problems and our next papers and lectures and grants…
    
    Given the state of the world in 2025, we should talk a bit more about how database research could help "fix“ the world. For instance, we can
        (a) help develop social networks and search engines not run and abused by tech oligarchs,
        (b) help make democracies resilient against fake news, lies, and anti-science,
        (c) help make our tech infrastructures resilient and independent from big companies, 
        (d) etc

    \item In our research community, the focus on performance improvements is a problem. And a {\bf huge} missed opportunity.

    \begin{center}
    {\it ``The best minds of my generation are thinking about how to make people click ads. That sucks.''}
    - Jeff Hammerbacher
    \vspace{.5em}
    
    {\it ``The best (database) minds of my generation are thinking about how to increase transaction throughput from one gazillion TAs/sec to 2 gazilllion TAs/sec. That sucks.''}
    - Jens Dittrich
         
    \end{center}
   
    \noindent How many people/companies in the world need this kind of insane performance? Could they all fit in a very small room?
    
    \item Instead, we should work more on other topics like usability, revise interfaces, make DBMSs usable for a wider crowd (in some unis a DB course is not even compulsary, there are quite a number of CS M.Sc.s who have no clue about databases). I believe part of the success of DuckDB is due to fixing usability issues in very nice ways.
    
    \item And, sure: keep on teaching the basic design patterns and success stories of our field (declarativity, physical data independence, automatic scaling) to the world (outside the realm of DBMSs), I mean, until AGI takes over…. (see 1.)
\end{myenumerate}

\section{A Sample of the Community}

We have collected a sample of positions from the broader data management community, attached below with minor grammatical editing.   

\subsection{Pınar Tözün}
I think there are "Systems for ML" aspects that our community is well-positioned to take on. For example, tackling data management challenges of deep learning such as data formats, IO path for deep learning, model and data caching, systematic benchmarking (not only for throughput/latency but also for sustainability), ...

First, I know the last one isn't unique to DB or ML, but we are a community that has a good tradition when it comes to creating standardized benchmarks and welcoming submissions focusing purely on experimental analysis and benchmarking in our venues. It is time to expand that tradition to incorporate resource-efficiency and sustainability in more fundamental ways rather than viewing them as nice-to-have add-ons. 

Second, for the large-scale applications, one may claim that the industry may be better positioned than academia to take on the challenge. However, we can always collaborate across industry and academia, and our graduates are appreciated by the industry. Furthermore, there are less commercial large-scale infrastructures (i.e., HPC) and smaller scale computing clusters (data centers at universities, resource-constrained setups, edge computing), where I see resource-aware and -efficient data management and processing is a must. Such infrastructures may have different needs than what is established in industry,  and our community can contribute in this direction.

\subsection{Leilani Battle}
I'm curious to explore not only our interests but also our ethics/responsibilities as data management researchers conversing with the broader field of "data science". Not just "what can we do?" but also "what should we do?"

In database research, we often see ourselves as humble workers toiling on some low-level abstraction layer that bears no responsibility for the transgressions of the abstractions above. However, my entire career has been built on the idea that we should have at least some idea of what is happening outside of the systems that we build. I take this a step further to say: we should consider not only the performance of external tools/applications and how their behavior may impact the performance of our own systems, but also whether these tools might use data management software in abusive or otherwise incorrect ways, and if possible, conduct new research to counteract potential external threats to the valid (and ideally, ethical) utilization of our work. Further, what does it mean for our community to develop ethical data processing and storage systems? I would love to see this become a more active topic of research for our community in the future.

\subsection{Aditya Parameswaran}

My view is that the database community needs to fully embrace LLMs as a means to solve the AI-complete problems that we care about, e.g., data integration, data cleaning, ... Plus, we also should help demonstrate how our principles of declarativity and query optimization can also help in LLM-powered processing at large/at scale/in production. At Berkeley, we've been having some luck with that for document/unstructured data processing, but there's a lot more we can do. Broadly, our community's focus on scale and efficiency is an asset, and we can't be the proverbial ostrich in the sand, but fully leverage, embrace, and contribute to the AI revolution

\subsection{Sudeepa Roy}

As researchers and more so in academia, we work on what we find interesting, important, and intellectually stimulating - this will always be there.  However, even as academic researchers, we have to serve the interests of our students, ensure acceptance of our research to the community, and get funding to support our research  - so our research will be shaped by whatever is trending, which currently is LLM. We also have to frequently answer questions from the reviewers of our database papers -  "could not LLM do it? Did you try ChatGPT”?. We have to find answers to these questions ourselves - whether or not reviewers ask. We as the database research community should pursue the passion of what we want to work on, but be open to embrace new ideas and take the opportunity to start new collaborations in the fast evolving landscape of large language models and other advances in ML/AI, related to the areas we care about as a community - query optimization and systems, data analysis, human-in-the-loop, database theory, database education in the classes we teach to train the next generation of experts on data, and everything else.

\subsection{Samuel Madden}
My personal perspective, which I think you know is that a) it’s not super healthy to wring our hands about the ways in which our community is missing out — we’ve been through this before with the web, mobile, “big data”, etc and we have continued to contribute as a community and be vibrant.  Looking over the people who are applying for faculty positions at MIT across systems and AI, many of them publish their interesting data systems papers in VLDB and SIGMOD.  I think that’s a strong sign that our community has a role to play in the new AI era.

\subsection{Felix Naumann}
The traditional problem space of all things data management is alive and kicking. Many (research) problems remain and have no chance of being eaten up by AI. Whoever has recently tried to install a DBMS, create a database and load a few simple CSV files into it knows firsthand: database systems are not the commodity we would like them to be. Whoever has collected and integrated several open data tables knows: more research is needed to make data accessible and useful. And anyone who has attempted to optimize complex queries across large-scale datasets discovers: the pursuit of efficient data processing remains an ongoing challenge.

\subsection{Paolo Papotti}

Over the past few years, I've shared aspects of this discussion with senior database experts in our community, particularly as I've observed that natural language processing (NLP) and deep learning specialists were more readily addressing "tabular data problems" with transformer-based solutions. One risk is the limited openness to new topics. A concrete example is the role of the key-value cache of LLMs and its connection to buffering to reduce inference time and cost. My experience is that papers from my group that are at the intersection of these topics have been well-received at TACL and NeurIPS, but we had mixed results with database venues.

While I see encouraging signs in "opening up" in the last year, with, for example, more topics added to our CFPs, I still believe that our community would benefit from putting a stronger emphasis on ideas when evaluating papers. In some discussions of papers under submission at DB venues, I notice that there is the risk of focusing more on the execution (the missing dataset, the extra baseline). This is done with the intention of improving papers, incorporating extensive technical content and expanding experimental evidence, but also with the risk that these aspects get more weight in the final decision compared to novelty and possible impact.

\subsection{James Cowling}
You might like our talk about SQL being a bad abstraction\footnote{\url{https://www.youtube.com/watch?v=dS9jtih4dI4}}. The OLTP database community feels very conservative to me compared to other branches of CS. We constantly hear "never bet against SQL" but I would certainly like more bets taking place.

The huge gift academia has is the ability to investigate \textbf{impractical ideas}. There's so much value to be unearthed by doing experimental things, challenging the status quo, and going in commercially non-viable directions in the search of found gold. All those Google papers got us thinking that research needs to be about commercially-viable ideas tested on production workloads, instead of a breeding ground for new and confronting ideas. No university is likely going to optimize an ads database better than a multi-trillion dollar company but they can help us see the world in new ways and birth new trillion dollar companies.

\balance

\bibliographystyle{ACM-Reference-Format}
\bibliography{main.bib}


\begin{thebibliography}{7}


\ifx \showCODEN    \undefined \def \showCODEN     #1{\unskip}     \fi
\ifx \showDOI      \undefined \def \showDOI       #1{#1}\fi
\ifx \showISBNx    \undefined \def \showISBNx     #1{\unskip}     \fi
\ifx \showISBNxiii \undefined \def \showISBNxiii  #1{\unskip}     \fi
\ifx \showISSN     \undefined \def \showISSN      #1{\unskip}     \fi
\ifx \showLCCN     \undefined \def \showLCCN      #1{\unskip}     \fi
\ifx \shownote     \undefined \def \shownote      #1{#1}          \fi
\ifx \showarticletitle \undefined \def \showarticletitle #1{#1}   \fi
\ifx \showURL      \undefined \def \showURL       {\relax}        \fi
\providecommand\bibfield[2]{#2}
\providecommand\bibinfo[2]{#2}
\providecommand\natexlab[1]{#1}
\providecommand\showeprint[2][]{arXiv:#2}

\bibitem[Abadi et~al\mbox{.}(2016)]%
        {abadi2016beckman}
\bibfield{author}{\bibinfo{person}{Daniel Abadi}, \bibinfo{person}{Rakesh Agrawal}, \bibinfo{person}{Anastasia Ailamaki}, \bibinfo{person}{Magdalena Balazinska}, \bibinfo{person}{Philip~A Bernstein}, \bibinfo{person}{Michael~J Carey}, \bibinfo{person}{Surajit Chaudhuri}, \bibinfo{person}{Jeffrey Dean}, \bibinfo{person}{AnHai Doan}, \bibinfo{person}{Michael~J Franklin}, {et~al\mbox{.}}} \bibinfo{year}{2016}\natexlab{}.
\newblock \showarticletitle{The Beckman report on database research}.
\newblock \bibinfo{journal}{\emph{Commun. ACM}} \bibinfo{volume}{59}, \bibinfo{number}{2} (\bibinfo{year}{2016}), \bibinfo{pages}{92--99}.
\newblock


\bibitem[Abadi et~al\mbox{.}(2020)]%
        {abadi2020seattle}
\bibfield{author}{\bibinfo{person}{Daniel Abadi}, \bibinfo{person}{Anastasia Ailamaki}, \bibinfo{person}{David Andersen}, \bibinfo{person}{Peter Bailis}, \bibinfo{person}{Magdalena Balazinska}, \bibinfo{person}{Philip Bernstein}, \bibinfo{person}{Peter Boncz}, \bibinfo{person}{Surajit Chaudhuri}, \bibinfo{person}{Alvin Cheung}, \bibinfo{person}{AnHai Doan}, {et~al\mbox{.}}} \bibinfo{year}{2020}\natexlab{}.
\newblock \showarticletitle{The Seattle report on database research}.
\newblock \bibinfo{journal}{\emph{ACM Sigmod Record}} \bibinfo{volume}{48}, \bibinfo{number}{4} (\bibinfo{year}{2020}), \bibinfo{pages}{44--53}.
\newblock


\bibitem[Bernstein et~al\mbox{.}(1998)]%
        {bernstein1998asilomar}
\bibfield{author}{\bibinfo{person}{Phil Bernstein}, \bibinfo{person}{Michael Brodie}, \bibinfo{person}{Stefano Ceri}, \bibinfo{person}{David DeWitt}, \bibinfo{person}{Mike Franklin}, \bibinfo{person}{Hector Garcia-Molina}, \bibinfo{person}{Jim Gray}, \bibinfo{person}{Jerry Held}, \bibinfo{person}{Joe Hellerstein}, \bibinfo{person}{HV Jagadish}, {et~al\mbox{.}}} \bibinfo{year}{1998}\natexlab{}.
\newblock \showarticletitle{The asilomar report on database research}.
\newblock \bibinfo{journal}{\emph{ACM Sigmod record}} \bibinfo{volume}{27}, \bibinfo{number}{4} (\bibinfo{year}{1998}), \bibinfo{pages}{74--80}.
\newblock


\bibitem[Dewitt(1995)]%
        {dewitt}
\bibfield{author}{\bibinfo{person}{J.~David Dewitt}.} \bibinfo{year}{1995}\natexlab{}.
\newblock \bibinfo{title}{Database Systems: Road Kill on the Information Superhighway?}
\newblock \bibinfo{howpublished}{VLDB Keynote}.
\newblock


\bibitem[Silberschatz and Zdonik(1996)]%
        {silberschatz1996strategic}
\bibfield{author}{\bibinfo{person}{Avi Silberschatz} {and} \bibinfo{person}{Stan Zdonik}.} \bibinfo{year}{1996}\natexlab{}.
\newblock \showarticletitle{Strategic directions in database systems—breaking out of the box}.
\newblock \bibinfo{journal}{\emph{ACM Computing Surveys (CSUR)}} \bibinfo{volume}{28}, \bibinfo{number}{4} (\bibinfo{year}{1996}), \bibinfo{pages}{764--778}.
\newblock


\bibitem[Stonebraker(2018)]%
        {stonebraker}
\bibfield{author}{\bibinfo{person}{Michael Stonebraker}.} \bibinfo{year}{2018}\natexlab{}.
\newblock \bibinfo{title}{My 10 Fears about the Future of the DBMS Field}.
\newblock \bibinfo{howpublished}{\url{https://www.jfsowa.com/ikl/Stonebraker.pdf}}.
\newblock


\bibitem[Wu(2024)]%
        {wublog}
\bibfield{author}{\bibinfo{person}{Eugene Wu}.} \bibinfo{year}{2024}\natexlab{}.
\newblock \bibinfo{title}{Where Does Database Research Go From Here?}
\newblock \bibinfo{howpublished}{\url{https://wp.sigmod.org/?p=3801}}.
\newblock


\end{thebibliography}

\end{document}